\documentclass[12pt,prl,superscriptaddress]{revtex4-1}
\usepackage{etoolbox}
\patchcmd{\section}
{\centering}
{\raggedright}
{}
{}
\patchcmd{\subsection}
{\centering}
{\raggedright}
{}
{}
\usepackage{amsmath,amssymb}
\usepackage{graphicx}
\usepackage{float}
\usepackage{subfigure}
\usepackage{verbatim}
\usepackage{lineno}
\usepackage{amsfonts}
\usepackage{dcolumn,tabularx}
\usepackage{bm}
\usepackage{color}
\usepackage[colorlinks,citecolor=blue]{hyperref}
\usepackage{times}
\usepackage{lineno}

\begin{document}

\title{Classical Communication Enhanced Quantum State Verification}
\author{Wen-Hao Zhang}
\author{Xiao Liu}
\author{Peng Yin}
\author{Xing-Xiang Peng}
\author{Gong-Chu Li}
\author{Xiao-Ye Xu}
\author{Shang Yu}
\author{Zhi-Bo Hou}
\author{Yong-Jian Han}
\author{Jin-Shi Xu}
\author{Zong-Quan Zhou}
\author{Geng Chen$\footnote{email:chengeng@ustc.edu.cn}$}
\author{Chuan-Feng Li$\footnote{email:cfli@ustc.edu.cn}$}
\author{Guang-Can Guo}
\affiliation{CAS Key Laboratory of Quantum Information, University of Science and Technology of China, Hefei, Anhui 230026, China.}
\affiliation{CAS Center For Excellence in Quantum Information and Quantum Physics, University of Science and Technology of China, Hefei, Anhui 230026, China.}
\date{\today}

\begin{abstract}
Quantum state verification provides an efficient approach to characterize the reliability of quantum devices for generating certain target states.
The figure of merit of a specific strategy is the estimated infidelity $\epsilon$ of the tested state to the target state, given a certain number of performed measurements $n$. Entangled measurements constitute the globally optimal strategy and achieve the scaling that $\epsilon$ is inversely proportional to $n$.
Recent advances show that it is possible to achieve the same scaling simply with non-adaptive local measurements, however, the performance is still worse than the globally optimal bound up to a constant factor.
In this work, by introducing classical communication, we experimentally implement an adaptive quantum state verification. The constant-factor is minimized from $\sim2.5$ to 1.5 in this experiment, which means that only 60\% measurements are required to achieve a certain value of $\epsilon$ compared to optimal non-adaptive local strategy. Our results indicate that classical communication significantly enhances the performance of quantum state verification, and leads to an efficiency that further approaches the globally optimal bound.

\end{abstract}

\maketitle
\section{INTRODUCTION}
\noindent Quantum information science aims to enhance traditional information techniques by introducing the advantage of ``quantumness". To date, the major subfields in quantum information include quantum computation \cite{Shor}, quantum cryptography \cite{Bennett}, and quantum metrology \cite{Giovannetti,Braunstein}, which are respectively in pursuit of more efficient computation, more secure communication, and more precise measurement. To achieve these innovations, one needs to manufacture quantum devices and verify that these devices indeed operate as expected. Various techniques have been developed for the task to inspect the quantum states generated from these devices. Quantum state tomography (QST) \cite{James} provides full information about an unknown state by reconstructing the density matrix and constitutes a popular point estimation method. However, the conventional tomographic reconstruction of a state is an exponentially time-consuming and computationally difficult process \cite{Hou}. In order to reduce the measurement complexity to certify the quantum states, substantial efforts have been made to formalizing more efficient methods. These improved methods normally require prior information or access partial knowledge about the states. On one hand, it has been found that with prior information about the category of the tested states, compressed sensing \cite{Flammia1,Gross} and matrix product state tomography \cite{Cramer} can be used to simplify the measurement of quantum states. On the other hand, entanglement witnesses can justify the appearance of entanglement with far fewer measurements \cite{Toth1,Toth2}; in a radical case, it is shown that local measurement on few copies is sufficient to certify the appearance of entanglement for multipartite entangled systems \cite{Dimic,Saggio}. Furthermore, when the applied measurements are correlated through classical communication, quantum tomography can be implemented in a significantly more efficient way \cite{Mahler,Qi,Chapman}.

In quantum information processing, the quantum device is generally designed to generate a specific target state. In this case, the user only needs to confirm that the actual state is sufficiently close to the target state, in the sense that the full knowledge about the exact form of the state is excessive for this requirement. Quantum state verification (QSV) provides an efficient solution applicable to this scenario.
As mentioned above, tomography aims to address the following question: What is the state? While QSV addresses a different question: Is the state identical/close to the target state? From a practical point of view, answering this question is sufficient for many quantum information applications. By performing a series of measurements on the output copies of state, QSV reaches a conclusion like ``the device outputs copies of a state that has at least $1-\epsilon$ fidelity with the target, with $1-\delta$ confidence".

In order to verify a specific quantum state, different kinds of strategies can be constructed; and thus, it is profitable for the user to seek an optimal strategy. Rigorously, this optimization can be achieved by minimizing the number of measurements of $n$ for given values of $\epsilon$ and $\delta$. Similar to the realm of quantum metrology \cite{Chen1,Chen2}, an optimal QSV strategy also strives for a $1/n$ scaling of $\epsilon$, with a minimum constant-factor before. For QSV, if the target state is a pure state, the best strategy is the projection onto the target state and its complementary space, then the $1/n$ scaling is reached, we call this strategy the globally optimal QSV strategy. Unfortunately, if the target is entangled state, entangled measurements are demanded while they are rare resources and difficult to obtain \cite{Gisin}. Recently, several works have shown that $1/n$ scaling can be achieved with a non-adaptive local (LO) strategy \cite{Pallister,Hayashi,Zhang4}, the LO here means that the applied measurement operators are separable as oppose to the entangled ones used in globally optimal strategy. However, this non-adaptive LO strategy is still worse than the globally optimal strategy by a constant-factor, which represents the number of additional measurements required to compete with the globally optimal strategy.

In this work, we demonstrate adaptive QSV using a photonic apparatus with active bi-directional feed-forward of classical communications between entangled photon pairs based on recent theoretical works\cite{Wang,Li,Yu}. The achieved efficiency not only attains the $1/n$ scaling but also further minimizes the constant-factor from before.
Both bi- and uni-directional classical communications are utilized in our experiment, and the results show that these adaptive strategies significantly outperform the non-adaptive LO strategy. Furthermore, the bi-directional strategy achieves higher efficiency than the uni-directional strategy, and the number of required measurements is reduced by $\sim40 \%$ compared to the non-adaptive LO strategy. Our results indicate that classical communication is beneficial resources in QSV, which enhances the performance to a level comparable with the globally optimal strategy.

\section{RESULTS}

\noindent Theoretical Framework

\noindent In a QSV task, the verifier is assigned to certify that his on-hand quantum device does produce a series of quantum states ($\sigma_1, \sigma_2, \sigma_3 ,...,\sigma_n$) satisfying the following inequality:
\begin{equation}
\label{prior information}
\langle\Psi| \sigma_i |\Psi\rangle >1-\epsilon \ (i=1,2,...,n).
\end{equation}
where $|\Psi\rangle$ is the target state that the device is supposed to produce. Eq. (\ref{prior information}) assumes a different scenario from that of QST, for which all $\sigma_i$ are required to be independent and identically distributed.

Typically, with the probability as $p_{l}(l=1,2,...,m) $, the verifier randomly performs a two-outcome local measurement $M_l$, which is accepted with certainty when performed on the target state. When all the measurement outcomes are accepted, the verifier can reach a statistical inference that the state from the tested device has a minimum fidelity $1-\epsilon$ to the target state, with a statistical confidence level of $1-\delta$.

For a specific strategy $\Omega=\Sigma_{l}p_{l}M_{l}$, the minimum number of measurements $n$ required to achieve certain values of $\epsilon$ and $\delta$ is then given by\cite{Wang}
\begin{equation}
\label{scaling}
n^{\Omega}_{local} =\frac{\ln(\delta)}{\ln[1-[1-\lambda^{\downarrow}_{2}(\Omega)]\epsilon]} \overset{\epsilon\rightarrow 0}{\approx}\frac{\ln(\delta^{-1})}{[1-\lambda^{\downarrow}_{2}(\Omega)]\epsilon}.
\end{equation}
This result indicates that it is possible to achieve the $1/n$ scaling of $\epsilon$ in the QSV of pure entangled states. Furthermore, the verifier can optimize the strategy by minimizing the second largest eigenvalue $\lambda^{\downarrow}_{2}(\Omega)$, as well as the constant-factor $\frac{1}{1-\lambda^{\downarrow}_{2}(\Omega)}$. For LO strategies with non-adaptive local measurements, the optimal strategy to verify $|\Psi(\theta)\rangle=\cos\theta|HV\rangle - \sin\theta|VH\rangle$ is identified with a minimum $\lambda^{\downarrow}_{2}(\Omega)$ as \cite{Pallister}
\begin{equation}
\label{lamda}
\lambda^{\downarrow}_{2}(\Omega_{LO}^{opt})=\frac{2+\sin2\theta}{4+\sin2\theta}
\end{equation}
The globally optimal strategy can be realized by projecting $\sigma_i$ to the target state $|\Psi\rangle$ and its orthogonal state $|\Psi^\perp\rangle$, under which $\lambda^{\downarrow}_{2}(\Omega)=0$; and thus, the globally optimal bound is calculated as
\begin{equation}
\label{global optimal}
n^{\text{opt}}_{\text{global}} =\frac{\ln(\delta^{-1})}{\epsilon}.
\end{equation}

For QSV of entangled states, entangled measurements are required to implement the globally optimal strategy, which are sophisticated to perform \cite{lutk,Vaidman,Calsamiglia,Ewert}. Therefore, local measurements are preferred from a practical view of point. This realistic contradiction naturally yields a question that how to further minimize the gap between locally and globally optimal strategies with currently accessible techniques.

Recently, a theoretical work generalizes the non-adaptive LO strategy to adaptive versions by introducing classical communication between the two parties sharing entanglement \cite{Wang}. The elementary adaptive strategy utilizes local measurements and uni-directional
classical communication (Uni-LOCC), as diagrammed in Fig. 1. The optimal Uni-LOCC QSV for $|\Psi(\theta)\rangle$ can be implemented by randomly choosing $M_{1}$, $ M_{2}$ or $M_{3}$ (see Methods for details) with prior probabilities $\{\frac{1}{2+2\sin^2\theta},\frac{1}{2+2\sin^2\theta},\frac{\sin^{2}\theta}{1+\sin^{2}\theta}\}$ ($\theta\in (45^\circ,90^\circ))$, and the corresponding strategy can be written as\cite{Wang}
\begin{equation}
\label{one-way optimal}
\Omega_{\rightarrow}=|\Psi(\theta)\rangle\langle\Psi(\theta)|+\frac{\sin^2\theta}{1+\sin^2\theta}
|\Psi^{\perp}(\theta)\rangle\langle\Psi^{\perp}(\theta)|+\frac{\cos^2\theta}{1+\sin^2\theta}|VV\rangle\langle VV|
+\frac{\sin^2\theta}{1+\sin^2\theta}|HH\rangle\langle HH|.
\end{equation}

A bi-directional LOCC (Bi-LOCC) strategy can be implemented by randomly switching the role between Alice and Bob, which can be denoted as $\Omega^{\rightarrow}_{\leftarrow}=|\Psi(\theta)\rangle\langle\Psi(\theta)|+
\frac{1}{3}(I-|\Psi(\theta)\rangle\langle\Psi(\theta)|)$.
Although both of these two strategies utilize one-step adaptive measurement, the Bi-LOCC strategy outperforms the Uni-LOCC when $\theta\neq45^{\circ}$.

When verifying entangled states with local measurements, adaptive strategies $\Omega_{\rightarrow}$ and $\Omega^{\rightarrow}_{\leftarrow}$ achieve higher efficiency compared to the non-adaptive LO strategy\cite{Wang,Yu}. The efficiency of LO, Uni-LOCC and Bi-LOCC strategies depend on their respective constant-factors $\frac{1}{1-\lambda^{\downarrow}_{2}(\Omega)}$, which are $2+\sin \theta \cos\theta$, $1+\sin^2 \theta$, and $3/2$. Although the performance of all these strategies coincides with two-qubit maximally entangled states ($\theta=45^{\circ}$), the adaptive strategies are still preferred in most practical scenarios, where the realistic states are always different from the maximally entangled ones and actually closer to target states with $\theta\neq45^{\circ}$.

\vspace {15pt}
\noindent Experimental implementation and results

\noindent In the above QSV proposals, a valid statement about the tested states is based on the fact that all the outcomes are accepted, while a single appearance of rejection will cease the verification without a quantified conclusion.
In practice, the generated states from the quantum devices are unavoidably non-ideal with a limited fidelity to the target state; and thus, there is always a certain probability to be rejected in each measurement. Even that the probability of single rejection is small, it is natural to observe rejection events in an experiment involving a sequence of measurements. As a result, these original proposals are likely to mistakenly characterize qualified quantum devices as unqualified, which is inadequate for experimental implementation.

By considering the proportion of accepted outcomes, a modified strategy is thus developed here, which is robust to a certain proportion of rejection events. Quantitatively, we have the corollary that if $\langle\Psi| \sigma_i |\Psi\rangle \leq 1-\epsilon$ for all the measured states, the probability for each outcome to be accepted is smaller than $1-(1-\lambda^{\downarrow}_{2}(\Omega))*\epsilon$. As a result, in the case that the verifier observes an accepted probability $ p\ge 1-(1-\lambda^{\downarrow}_{2}(\Omega))*\epsilon$, it should be concluded that the actual state satisfies Eq. (\ref{prior information}) with a confidence level of $1-\delta$, where $\epsilon$ and $\delta$ are calculated from the inequality \cite{Dimic}
\begin{equation}
\label{tensorproduct}
\delta \leq e^{-D(\frac{m}{n}||1-(1-\lambda^{\downarrow}_{2}(\Omega))\epsilon) n},
\end{equation}
with
\begin{equation}
\label{div}
\begin{split}
&D(x||y)=x \log \frac{x}{y}+(1-x) \log \frac{1-x}{1-y}, \\
\end{split}
\end{equation}
 and $m$ results are accepted when $n$ measurements are performed. As a result of this modification, in the case that the final accepted probability $ p\ge 1-(1-\lambda^{\downarrow}_{2}(\Omega))*\epsilon$, the verification can eventually reach a conclusion quantifying the distance between the actual and target states.

Benefiting from this modification, QSV can be applied to realistic non-ideal states, which allows us to experimentally verify two-qubit entangled states using the above adaptive proposals. With the setup shown in Fig. \ref{setup}, we can perform adaptive QSV. The setup consists of an entangled photon-pair source (see Methods for details), two mechanical optical-switcher (MOS), and two high-speed triggered polarization analyzer (TPA). For adaptive QSV, Alice can guide her photon towards the MOS and perform a randomly selected projective measurement by TPA. Afterward, through a unidirectional classical communication, Alice's outcome is sent to Bob to control the measurement performed on the paring photon, which is delayed on Bob's MOS. An opposite adaptive process can also be realized by switching the role of Alice and Bob; and thus, the symmetric adaptive QSV can be executed by randomly selecting the two communication directions with equal probabilities. Technically, this random adaptive operation can be realized by controlling the MOS with a quantum random number generator (QRG), which outputs a binary signal (0 and 1) to decide which MOS transmits the photon directly while the other MOS delays the passing photon. For both Uni- and Bi-LOCC strategies, we use a QRNG to randomly decide the applied setting among ${M_{1}, M_{2}, M_{3}}$; therefore, the settings are unknown to the incident photon pairs in prior to the measurement.

In order to confirm the power of classical communication in QSV, three strategies (LO, Uni-LOCC, and Bi-LOCC) are utilized to verify a partially entangled state $|\Psi(60^{\circ})\rangle$ and the results are shown in Fig. \ref{60}. In Fig. \ref{60} (a), the results of 50 trials are averaged, which approximately coincide with the theoretical lines for the first few measurements and deviate from the predicted linearity afterward. This deviation mainly results from the difference between verified states and the ideal target state, which leads to rejection outcomes in QSV. In other words, only if the verified states are perfectly identical to the target state, a persistent $1/n$ scaling can be observed in a practical QSV. Since the occurrence rates of rejections are in principle equal for different strategies, a distinct gap in the estimated fidelity can be seen between the adaptive and non-adaptive strategies as predicted in the theory part. These results indicate the power of classical communication in boosting the performance of QSV. However, the practical scaling is not only determined the optimality of the strategy, but also the quality of the actual state. In this sense, we can only access the intrinsic performance of a strategy by testing an ideal state. Although it is impossible to generate an ideal state in experiment, we can circumvent this difficulty by studying the first few measurements, of which the occurrence of rejections is fairly rare.
In Fig. \ref{60} (b), the first 25 measurements of single trials with all the outputs to be accepted are plotted, accompanied by the averaged results in Fig. \ref{60} (a) in the same range. The efficiency can be characterized by the slope of linear fitting lines of these data points. For LO, Uni-LOCC, and Bi-LOCC strategies, the fitted slope values of the averaged points are 0.13, 0.17, and 0.187, respectively. After eliminating the effects of the state deviations by considering all-accepted single trials, the fitted slope values are 0.135, 0.188, and 0.22 for LO, Uni-LOCC, and Bi-LOCC strategies respectively, and these values are exactly the theoretical predictions for ideal states. As a result, the efficiency of the Bi-LOCC strategy is 1.63 times higher than that of LO and 1.17 times higher than that of Uni-LOCC. In other words, by introducing bi-directional classical communications, only $~60\%$ measurements of the non-adaptive LO scenario are required to verify the states to a certain level of fidelity. The performance gap between optimal local strategy and the globally optimal strategy is further minimized. Concretely, the constant-factor $\frac{1}{1-\lambda^{\downarrow}_{2}(\Omega)}$ is reduced to approximately 1.5.

A further study of the performance gap between Uni- and Bi-LOCC strategies is made to verify another two entangled states $|\Psi(70^{\circ})\rangle$ and $|\Psi(80^{\circ})\rangle$, and the averaged results of 50 trials are shown in Fig. \ref{70&80}. Both results show that Bi-LOCC significantly outperforms Uni-LOCC, and the differences of estimated fidelities are 2.1\% and 1.3\% for $|\Psi(70^{\circ})\rangle$ and $|\Psi(80^{\circ})\rangle$, respectively. Classical communication better enhances QSV by transferring the information bi-directionally rather than an ordinary uni-directional configuration.

\section{Discussion}
\noindent One main motivation to explore the quantum resources, such as entangled states and measurements, is their potential power to surpass the classical approaches. On the other hand, the fact that the quantum resources are generally complicated to produce and control inspires another interesting question: how to use classical resources exhaustively to approach the bound set by quantum resources? In the task to verify an entangled state, the utilization of entangled measurements constitutes a globally optimal strategy that achieves the best possible efficiency. Surprisingly, one can also construct strategies merely with local measurements and achieve the same scaling. In this experiment, we show that by introducing classical communications into QSV, the performance with local measurements can be further enhanced to approach the globally optimal bound. As a result, to verify the states to a certain level of fidelity, the number of required measurements is only 60\% of that for non-adaptive local strategy. Meanwhile, the gap between the locally and globally optimal bound is distinctly reduced, with the constant-factor minimized to ~1.5 before $1/n$ scaling. Furthermore, recently QSV has been generalized to the adversarial scenario where arbitrary correlated or entangled state preparation is allowed\cite{Zhu2019,Zhu2019pra}.

\section{Methods}
\noindent Generation of entangled photon pairs.

\noindent In the first part of the setup, tunable two-qubit entangled states are prepared by pumping a nonlinear crystal placed into a phase-stable Sagnac interferometer (SI). Concretely, a 405.4 nm single-mode laser is used to pump a 5mm long bulk type-II nonlinear periodically poled potassium titanyl phosphate (PPKTP) nonlinear crystal placed into a phase-stable SI to produce polarization-entangled photon pairs at 810.8 nm. A PBS followed by an HWP and a PCP are used to control the polarization mode of the pump beam. These lenses before and after the SI are used to focus the pump light and collimate the entangled photons, respectively. The interferometer is composed of two highly reflective and polarization-maintaining mirrors, a Di-HWP and a Di-PBS. ``Di" here means it works for both 405.4 nm and 810.8 nm. The Di-HWP flips the polarization of passing photons, such that the type-II PPKTP can be pumped by the same horizontal light from both clockwise and counterclockwise directions. Di-IF and LPF (Long pass filter) are used to remove the pump beam light. BPF (bandpass filter) and SMF are used for spectral and spatial filtering, which can significantly increase the fidelity of entangled states.
The whole setup, in particular the PPKTP, is sensitive to temperature fluctuations. Placing the PPKTP on a temperature controller ($\pm0.002^\circ $C stability) and sealing the SI with an acrylic box would help improve temperature stability. Polarization-entangled photon pairs are generated in the state $|\Psi(\theta)\rangle=\cos\theta|HV\rangle - \sin\theta|VH\rangle$ ($H$ and $V$ denote the horizontally and vertically polarized components, respectively) and $\theta$ is controlled by the pumping polarization.
\vspace {15pt}

\noindent Measurement setting for adaptive QSV

\noindent For the QSV of two-qubit pure entangled states, Alice's measurement $\Pi_{i}$ (i=1,2,3) are selected to be Pauli X,Y and Z measurements. When the outcome of X, Y, and Z is 1(0), Bob performs $\Pi_{11}=|\upsilon^{+}\rangle\langle\upsilon^{+}|$ ($\Pi_{10}=|\upsilon^{-}\rangle\langle\upsilon^{-}|$), $\Pi_{21}=|\omega^{+}\rangle\langle\omega^{+}|$ ($\Pi_{20}=|\omega^{-}\rangle\langle\omega^{-}|$) and $\Pi_{31}=|V\rangle\langle V|$ ($\Pi_{30}=|H\rangle\langle H|$), respectively, and the vectors are defined as $|\upsilon^{\pm}\rangle=\sin\theta|H\rangle\mp\cos\theta|V\rangle $ and $|\omega^{\pm}\rangle=\sin\theta|H\rangle\pm i\cos\theta|V\rangle $. These adaptive measurement settings constitute the optimal Uni-LQCC strategy which has the form \cite{Wang}
\begin{equation}
\label{Uni elements}
\begin{split}
&M_{1}=|+\rangle\langle+|\otimes|\upsilon^{+}\rangle\langle\upsilon^{+}|
+|-\rangle\langle-|\otimes|\upsilon^{-}\rangle\langle\upsilon^{-}|,\\
&M_{2}=|R\rangle\langle R|\otimes|\omega^{-}\rangle\langle\omega^{-}|
+|L\rangle\langle L|\otimes|\omega^{+}\rangle\langle\omega^{+}|,\\
&M_{3}=|V\rangle\langle V|\otimes|H\rangle\langle H|
+|H\rangle\langle H|\otimes|V\rangle\langle V|,\\
\end{split}
\end{equation}
where $|+\rangle\equiv\frac{1}{\sqrt{2}}(|H\rangle+|V\rangle)$ and $|-\rangle\equiv\frac{1}{\sqrt{2}}(|H\rangle-|V\rangle)$ denote the eigenstates of Pauli X operator, $|R\rangle\equiv\frac{1}{\sqrt{2}}(|H\rangle+i|V\rangle)$ and $|L\rangle\equiv\frac{1}{\sqrt{2}}(|H\rangle-i|V\rangle)$ denote the eigenstates of Pauli Y operator.
In each of these three combined local measurement settings, the choice of Bob's measurement setting is determined by the outcome of Alice's measurement, which can be achieved by controlling the local operation of Bob's EOM according to Alice's outcome.

\section{DATA AVAILABILITY}
\noindent The authors declare that all data supporting the findings of this study are available within the article or from the corresponding author upon reasonable request.

\section{ACKNOWLEDGEMENTS}
\noindent This work was supported by the National Key Research and Development Program of China (Nos. 2016YFA0302700, 2017YFA0304100), National Natural Science Foundation of China (Grant Nos. 11874344, 61835004, 61327901, 11774335, 91536219, 11821404), Key Research Program of Frontier Sciences, CAS (No. QYZDY-SSW-SLH003), Anhui Initiative in Quantum Information Technologies (AHY020100, AHY060300), the Fundamental Research Funds for the Central Universities (Grant No. WK2030020019, WK2470000026), Science Foundation of the CAS (No. ZDRW-XH-2019-1).

\section{AUTHOR CONTRIBUTIONS}
\noindent W.-H.Z. made the calculations assisted by P.Y.and J.-S.X. C.-F.L. and G.C. planned and designed the experiment. W.-H.Z. carried out the experiment assisted by G.C., X.L., G.-C.L., X.-Y.X., S.Y., Z.-B.H., Y.-J.H., and Z.-Q.Z. whereas W.-H.Z. and X.-X.P.. designed the computer programs. W.-H.Z. and G.C. analyzed the experimental results and wrote the manuscript. G.-C.G. and C-F.L. supervised the project. All authors discussed the experimental procedures and results.

\section{ADDITIONAL INFORMATION}
\noindent \textbf{Competing interests:} The authors
declare that they have no competing interests.

\section{REFERENCES}
{}

\begin{figure}[htbp]
\centering
\includegraphics[width=6in]{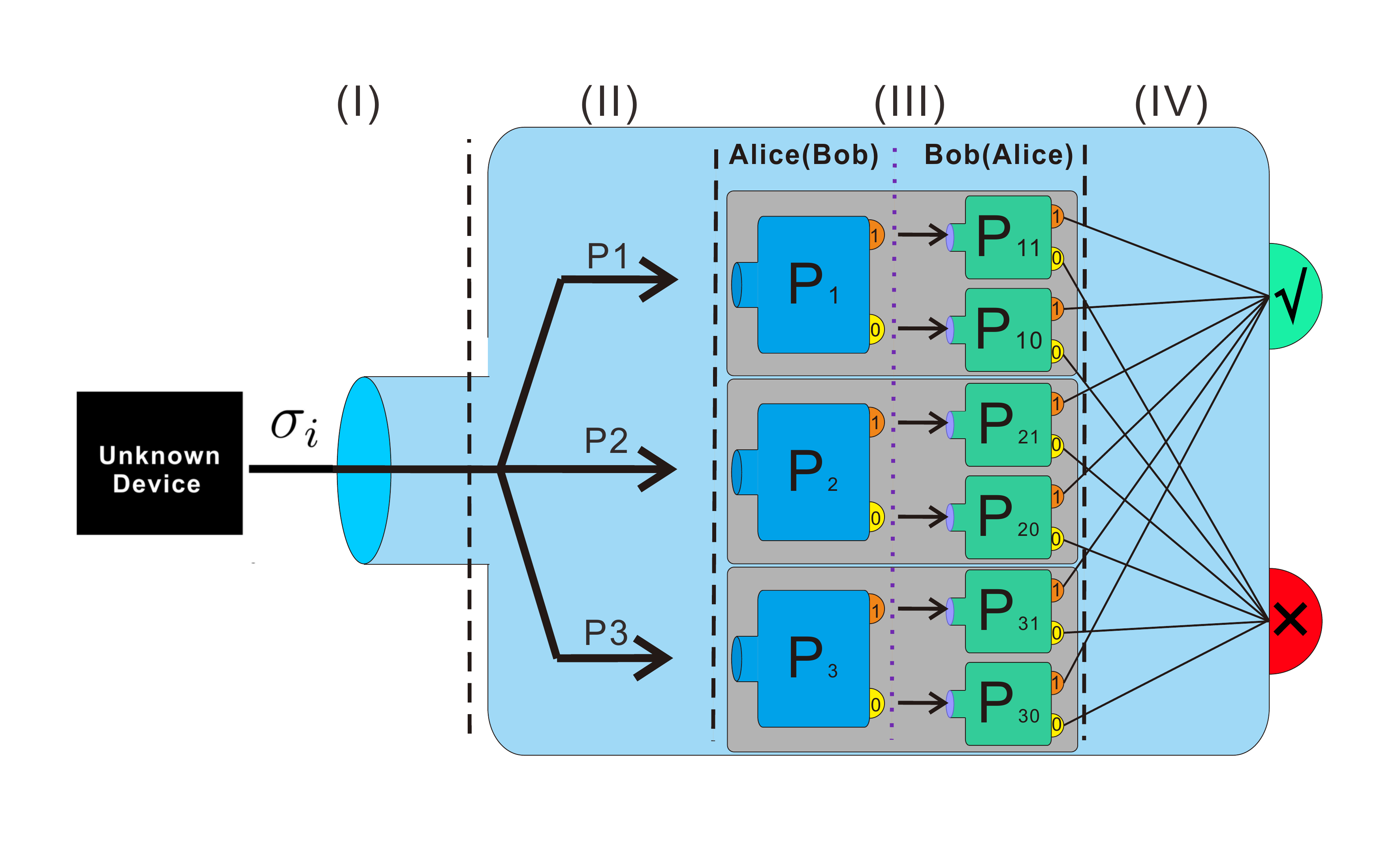}
\caption{Diagram of adaptive QSV with LOCC. The figure represents the general procedure to implement an adaptive QSV to verify whether an unknown source generates a pure target state. The generated states $\sigma_1, \sigma_2, \sigma_3 ,...,\sigma_n$, are projected firstly by Alice with a randomly selected measurement setting $\Pi_{i}$, with a prior probability as $P_{i}$. The measurement on Bob's side depends on the outcome of Alice's measurement, i.e., if the outcome of $\Pi_{i}$ is 0 (1), Bob performs measurement $\Pi_{i0}$ ($\Pi_{i1}$) accordingly. Bob's outcome 1 and 0 are coarse-grained as accepted ($\surd$) and rejected ($\times$) events, respectively. Similarly, Alice can perform measurements according to Bob's outcome. A bi-directional strategy can be applied by performing these two uni-directional strategies randomly. Through a statistical analysis of the sequence of accepted and rejected events, the verifier can ascertain the largest possible distance between the actual and target states up to some finite statistical confidence.}
\label{diagramm}
\end{figure}

\begin{figure}[htbp]
\centering
\includegraphics[width=5in]{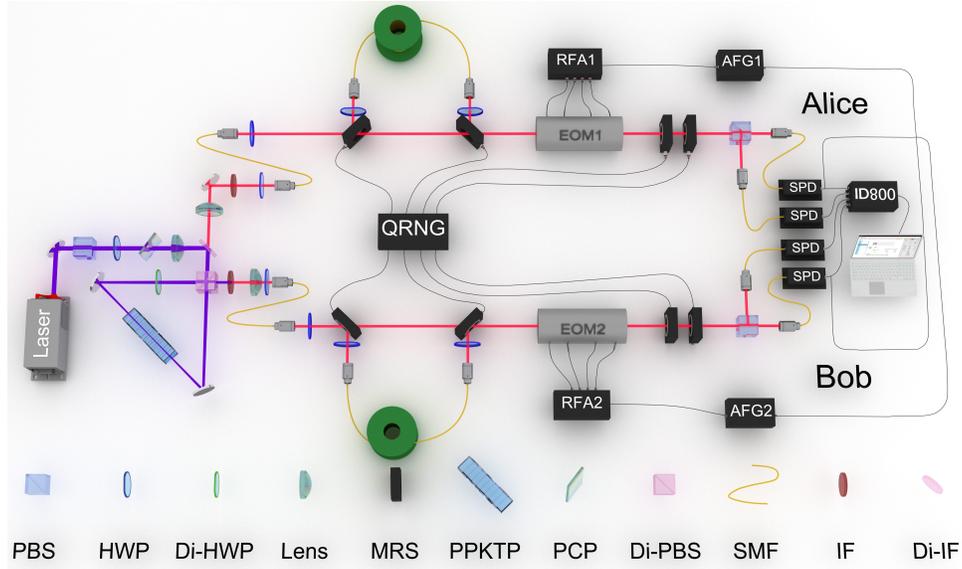}
\caption{Experimental setup. The setup includes an entanglement source, two sets of MOS, and two sets of TPA. The entanglement source is mainly a SI in a triangle configuration, and the generated photon pairs are distributed to two separate parties, namely, Alice and Bob. The MOS consists of two D-Shaped Mirrors mounted in a motorized rotation stage (MRS), and a 100-meter single-mode fiber (SMF) for delaying. Each TPA is composed of one electro-optical modulator (EOM) and one standard polarization analyzers, which consists of one half-wave plate (HWP) and one quarter-wave plate (QWP) mounted in MRSes, and a following polarized beam splitter (PBS) with two single-photon detectors (SPDs) at its two exits. For the Uni-LQCC protocol, the MOS on Alice's side is rotated to be open for the passing photon, which is directly measured by TPA with a randomly selected measurement setting. The other photon on Bob's side is reflected in the SMF at Bob's MOS, and then reflected into the TPA after emitting from the SMF. Adaptive processing is realized by using the outcome of Alice's measurement to control the measurement setting of Bob's TPA. Technically, Alice's photon sparks the corresponding SPD and the produced pulse triggers an arbitrary function generator (AFG) to output a recognizable signal for the EOM. After amplifying this signal to an adequate amplitude by a radio-frequency amplifier (RFA), the EOM can be driven to perform a required operation on the passing photon; and thus, Bob's measurement is adaptive to the outcomes of Alice's measurement. The coincidence is recorded and analyzed by an ID800 (ID Quantique). The Bi-LQCC protocol can be implemented by randomly switching the role of Alice and Bob. The randomness of the measurement setting and communication direction is realized by controlling the TPA and MOS with the QRNG, respectively.
Di - dichroic, MMF - multi-mode fiber, IF - interference filter, SMF - single-mode fiber, PCP - phase compensation plate, L - lens. }
\label{setup}
\end{figure}

\begin{figure}[htbp]
\centering
\includegraphics[width=6in]{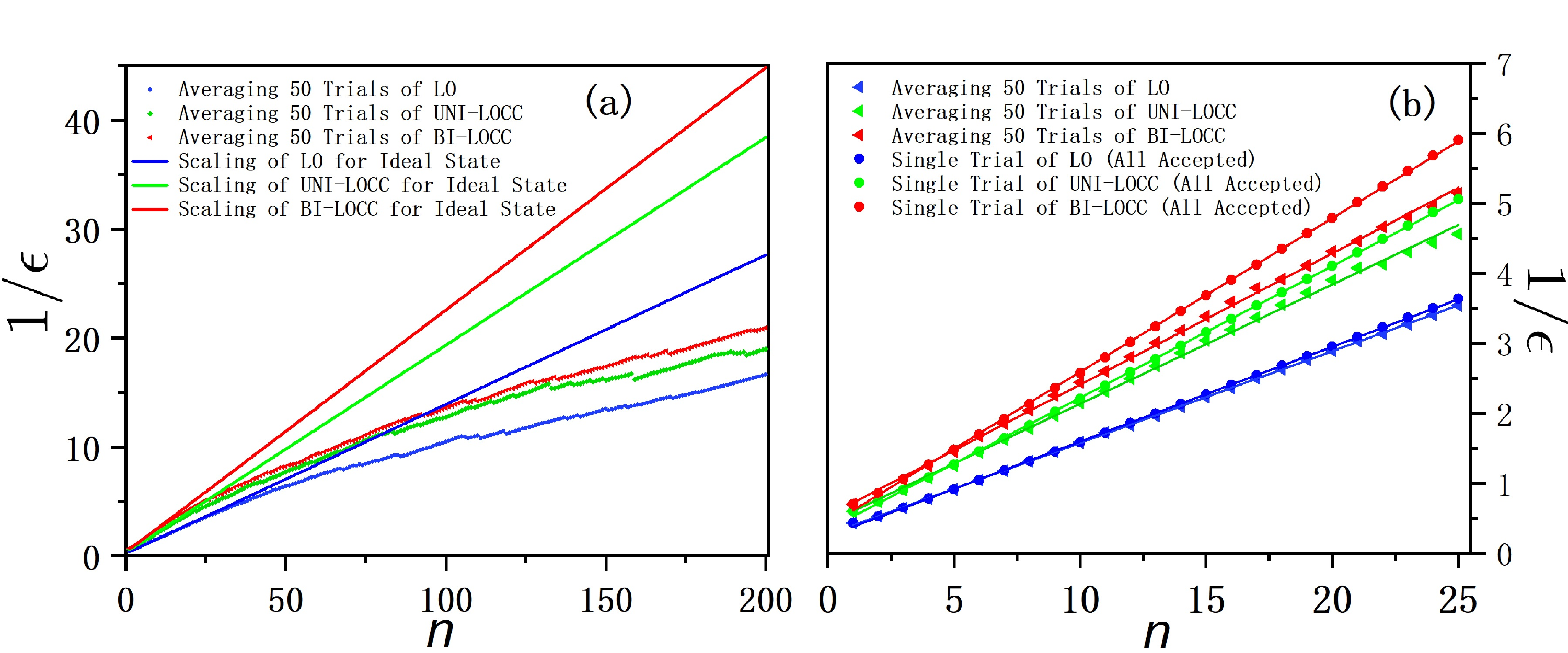}
\caption{Experimental QSV of partially entangled state $|\Psi(60^{\circ})\rangle$. Three strategies (LO, Uni-LOCC, and Bi-LOCC) are utilized in this experiment to verify the entangled state $|\Psi(60^{\circ})\rangle$. For each strategy, a total of 50 trials of QSV are performed, and each trial contains 200 measurements. In both (a) and (b), $\delta$ is set to 0.05, and the value of $1/\epsilon$ is plotted against the number of measurement $n$. The averaging results of the 50 trials are shown in (a) together with the theoretical lines for the ideal state. The averaging results suffer from the decoherence and noise of the realistic states and measurements, and thus, deviate from the predicted linearity gradually with $n$. To learn the true performance of the applied strategies, the results of single trials in which all the first 25 measurements output acceptance are shown in (b), together with the averaging results of 50 trials in the same range. For each case in (b), the data points are linearly fitted and the performance can be quantitatively characterized by the slope of the fitting line. The standard deviation for the averaging data points approximately grows linearly with $n$ and is expressed as $\varepsilon=sn$, with $s$ equals to 0.039, 0.052, and 0.049 for the LO, Uni-LOCC, and Bi-LOCC strategies, respectively.}
\label{60}
\end{figure}

\begin{figure}[htbp]
\centering
\includegraphics[width=6in]{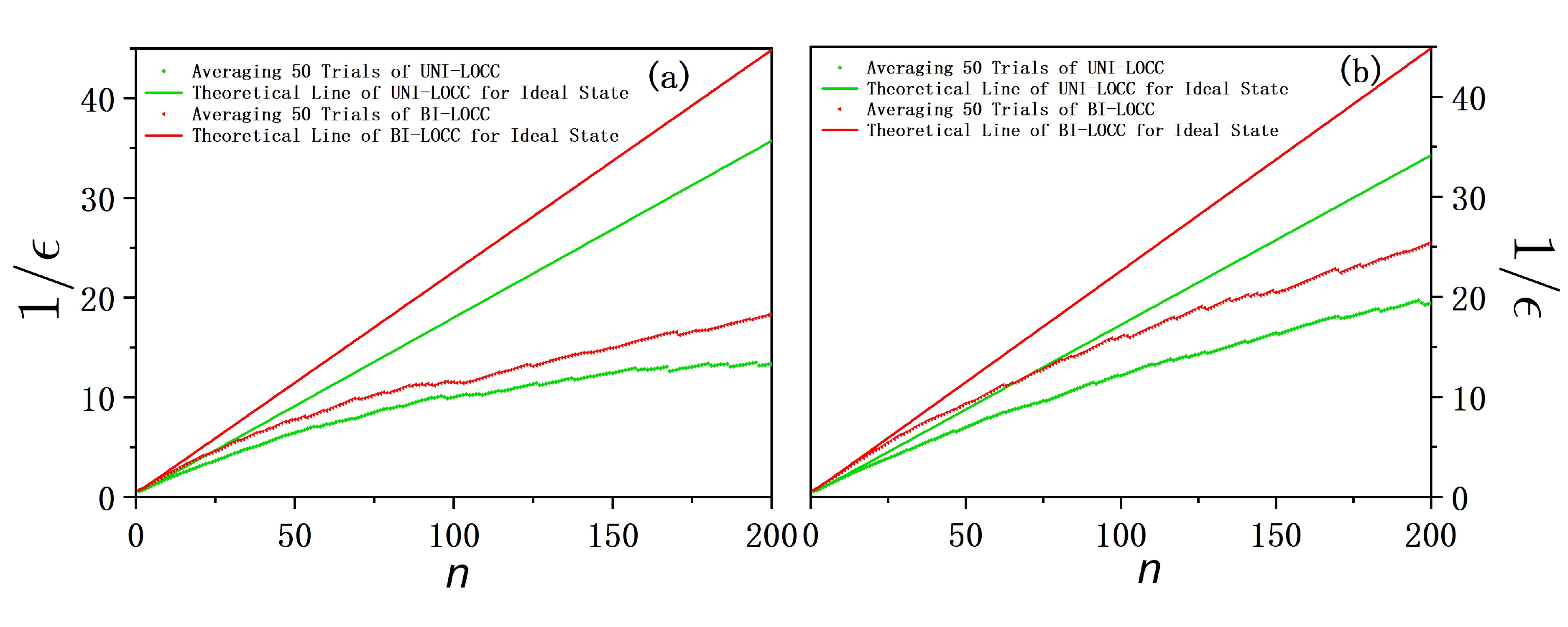}
\caption{Experimental QSV of partially entangled states $|\Psi(70^{\circ})\rangle$ and $|\Psi(80^{\circ})\rangle$. Two adaptive strategies, Uni-LOCC and Bi-LOCC are utilized to study the performance gap between them. In both (a) and (b), $\delta$ is set to 0.05, and the value of $1/\epsilon$ is plotted against the number of measurement $n$. The averaged results of 50 trials for $|\Psi(70^{\circ})\rangle$ and $|\Psi(80^{\circ})\rangle$ are shown together with the theoretical lines for ideal states. For both verified states, a distinct gap in the estimated fidelity can be seen between the Uni- and Bi-LOCC strategies. The standard errors for the averaging data points can be approximately expressed as $\varepsilon=sn$ with $s= 0.037$ and 0.035 for UNI- and BI-LOCC strategies in (a), and $s= 0.045$ and 0.057 for UNI- and BI-LOCC strategies in (b).}
\label{70&80}
\end{figure}

\end{document}